\documentclass[prl,showpacs,twocolumn]{revtex4}
\usepackage{graphicx}
\usepackage{bm}

\begin{document}

  \title{Response of Complex Systems to Complex Perturbations: the Complexity Matching Effect}

\author{Paolo Allegrini$^{1}$}
\author{Mauro Bologna$^{2}$}
\author{Paolo Grigolini$^{3,4,5}$}
\author{Bruce J. West$^{1,6}$}

\affiliation{$^{1}$Physics Department, Duke University, Durham NC 27708 USA.\\
$^2$Dep. de F\'{i}sica,
Universidad de Tarapac\'{a}, Campus Vel\'{a}squez, Vel\'{a}squez
1775, Casilla 7-D, Arica, Chile.\\
$^3$Center for Nonlinear Science, University of North Texas,
    P.O.Box 311427, Denton TX 76203-1427 USA.\\
$^{4}$Dipartimento di Fisica "E. Fermi"-Universit\`a di Pisa and
INFM, Largo Pontecorvo 3, 56127 Pisa, Italy.\\
$^{5}$Istituto dei Processi Chimico Fisici del CNR
Area della Ricerca di Pisa, Via G. Moruzzi 1,
56124 Pisa, Italy.\\
$^6$Army Research Office,  Research Triangle Park, NC 27709 USA.}
\date{\today}

\begin{abstract}
The dynamical emergence (and subsequent intermittent breakdown) of collective behavior in
complex systems is described as a non-Poisson renewal process, characterized
by a waiting-time distribution density $\psi (\tau )$ for the time intervals
between successively recorded breakdowns. In the intermittent case $\psi (t)\sim
t^{-\mu }$, with complexity index $\mu $. We show that two systems can
exchange information through complexity matching and present theoretical and
numerical calculations describing a system with complexity index $\mu _{S}$
perturbed by a signal with complexity index $\mu _{P}$. The analysis focuses
on the non-ergodic (non-stationary) case $\mu \leq 2$ showing that for $\mu _{S}\geq \mu _{P}$, 
the system $S$ statistically inherits the correlation function of the
perturbation $P$. The condition $\mu _{P}=\mu _{S}$ is a resonant maximum
for correlation information exchange. 
\end{abstract}

\maketitle




 The response of complex systems to external excitations has been one of the central issues of statistical physics for the past fifty years. These analyses started with the work of Kubo \cite{green} and adopted the assumption that the unperturbed system can be described by a correlation function with a finite correlation time. Stochastic resonance (SR) is one phenomenon that, although consistent with the assumptions of statistical physics, yields counter-intuitive results \cite{vulpiani} and has been a popular topic of investigation for the past quarter century. SR is a strategy for the transmission of information through random (Poisson statistics) media that can enhance the signal-to-noise ratio \cite{stochasticresonance}. Another popular topic of investigation has been synchronization of the dynamic elements in a complex network, whether the elements are modeled as linear \cite{kuramoto} or nonlinear oscillators \cite{winfree}, limit cycles, or chaotic attractors \cite{arkady1}. Both the mechanisms of synchronization and SR have been used to provide insight into the behavior of complex networks as, for example, into the dynamics of collections of neurons \cite{moss}. Even more recently the question of how to transmit information through a complex network in which the simplifying assumptions of traditional statistical physics are no longer valid has been addressed \cite{hanggi,barbi}. 
 Barbi {\em et al.}  conclude that a complex network described by intermittent fluctuations with non-Poisson statistics does not respond to external periodic perturbations \cite{barbi} in the non-stationary (or non-ergodic) regime. Other investigators 
 have reached the same conclusion \cite{recentobservation}
 
 A number of complex phenomena have been shown to have non-Poisson statistical properties, including collections of blinking quantum dots \cite{bqd} and collections of neurons in the human brain \cite{simone}. The non-Poisson character of the distribution of the sojourn times in the ``on" and ``off" states in BQDs is well known \cite{bqd}, however the renewal character of the statistics has only recently been established \cite{brokman}. Moreover, Bianco {\em et al.} \cite{massi} have shown using a network of coupled two-state stochastic clocks that with the onset of phase synchronization at a critical value of the coupling coefficient the dynamics of the network becomes that of a non-Poisson renewal process operating in the non-ergodic regime. 
The breakdowns of the collective phase structures
reset 
 memory of previous state changes, consequently yielding a non-Poisson renewal process (NRP). 
 NPRs are characterized by a waiting-time distribution density between events (collisions), indicated with $\psi(\tau)$. The regression to equilibrium of an ensemble of NPRs is given by the survival probability $\Psi(t)=\int_0^{\tau}\psi(\tau)d\tau$. In this letter we focus our attention on the case
 $\psi(\tau)\approx \tau^{-\mu}$, with $1\leq\ \mu <2$, yielding inverse power-law relaxation of the form $\Psi(t)\approx t^{-\nu}$ with index $\nu<1$, thereby violating the finite time-scale assumption, so often made. One way a non-exponential decay of this kind can be expressed is as the sum of infinitely many Poisson components, but if these components are independent, as are the single units in the absence of cooperation, no NPR events are generated \cite{barbino}. The production of NPR events is a sign of close cooperation among distinct units, thereby offering a rational as to why NPR processes do not respond to periodic perturbations \cite{barbi,recentobservation}, as would single independent Poisson units. NPR processes reflect a condition shared by the phenomenological models of glassy dynamics \cite{glassy}, laser cooling \cite{book} and models of atomic transport in optical lattices \cite{atomic}. Other NPR processes have been found at the core of the correlations in DNA sequences \cite{indian}, heart-beat variability \cite{heart} and earthquakes \cite{mega}.

 Herein we extend the investigation of Refs. \cite{barbi,recentobservation} to determine how one complex network responds to a perturbation by a second complex network as a function of the matching of the measures of complexity of the two networks. Here the complex system is an NPR network and the measure of complexity is the inverse power-law index. More precisely, we consider a NPR system, with power-law index $\mu<2$, and study, as done in the theory of SR \cite{stochasticresonance}, the case where the rate of production of jumps, a kind of renewal event, is modulated by an external perturbation. However, we do not use the harmonic perturbation of \cite{barbi,recentobservation}, but instead use a random signal as perturbation. This is to some extent reminiscent of the condition of aperiodic SR \cite{aperiodic}, the significant differences being that the network satisfies the NPR condition with $\mu_S<2$ and the perturbation is another NPR process with power index $\mu_P<2$. We prove that the NPR network is sensitive to the influence of the NPR perturbation and that the response intensity becomes maximum at the matching condition $\mu_S=\mu_P$. 
We refer to this as the Complexity Matching Effect (CME), insofar as within our theoretical framework, genuinely complex processes are NPR systems, fitting the non-stationary condition $\mu<2$: the parameter $\mu$ is the complexity index and two systems with the same $\mu$ share the same degree of complexity.

Although NPR processes are 
widespread in the complexity field, 
we adopt the 
BQD
picture \cite{bqd}
to illustrate the important concepts of jump and collision. The intermittent
BQD fluorescence is a sequence of \emph{jumps} from the ``light on'' to the
``light off'' state and vice-versa. We use this sequence to generate the
dichotomous variable $\xi _{S}(t)$, with the values +$1$ and $-1$
corresponding to the system in the ``light on'' (or $|+\rangle $) and
``light off'' (or $|-\rangle $) state, respectively. These jumps are events
whose renewal nature has been proved with compelling arguments by the
statistical analysis in Refs. \cite{brokman}. Furthermore, the time interval
between consecutive jumps is described by a histogram with the form of
an inverse power law $\psi ^{(exp)}(\tau )\sim \tau ^{-\mu }$ with power
index $\mu <2$. Thus the fluctuation $\xi _{S}(t)$ is a NPR process, which,
according to Bel and Barkai \cite{bel}, violates the ergodic condition. For
simplicity's sake we assume that the distribution densities of sojourn times
in the two states are identical.
The relaxation process from an initial out of equilibrium condition is
described by 
\begin{equation}  \label{correlationfunction}
\Psi (t) = \left [\frac{T}{t + T} \right ]^{\mu-1},
\end{equation}
therefore with 
$\nu = \mu -1$. 

This relaxation process is not directly related to the distribution density
of time intervals between two consecutive jumps, denoted by $\psi
^{(exp)}(\tau )$, but to another distribution density, indicated by the
symbol $\psi (\tau )$, namely the waiting-time distribution for \emph{%
collisions}. 
A collision is a non-Poisson event as renewal as a jump, associated with a
coin tossing that selects the sign of the state produced by the collision
event. In the absence of perturbation, the probability of selecting the
state $|+\rangle $ is identical to the probability of selecting the state $%
|-\rangle $. Consequently, the time interval between two consecutive
collisions, with distribution density $\psi (\tau )$, can be shorter than
the time interval between two consecutive jumps, but the two distributions, $%
\psi (\tau )$ and $\psi ^{(exp)}(\tau )$, have the same power-law index $\mu
<2$ \cite{gerardo}. The choice of the form of Eq. (\ref{correlationfunction}%
) generates for $\psi (\tau )$ the analytical form 
\begin{equation}
\psi (\tau )=(\mu -1)\frac{T^{\mu -1}}{(\tau +T)^{\mu }}.
\label{theoretical}
\end{equation}

Note that the relaxation function of Eq. (\ref{correlationfunction}),
playing the role of correlation function, is the survival probability
associated with the waiting time distribution of Eq. (\ref{theoretical}).
Note also that in general the correlation function of a NPR is not
stationary and ``ages'' with time, namely, with the time lag between
preparation and observation. In fact, the probability density of the first
waiting time $\tau _{1}$ depends on the time lag between preparation and
observation. When the observation begins at the same time as preparation,
the system is said to be ``young'' and the p.d.f. of $\tau _{1}$ coincides
with $\psi (\tau )$ \cite{gerardo}. The corresponding survival probability,
Eq. (\ref{correlationfunction}), is the young correlation function. The
distribution density of Eq. (\ref{theoretical}) is normalized, and the
parameter $T$, keeping the distribution density finite also for $\tau
\rightarrow 0$, gives information on the lapse of time necessary to reach
the time asymptotic condition at which $\psi _{S}(\tau )$ becomes identical
to an inverse power law. The theoretical results
of this Letter are asymptotic in time, and do not depend on the specific
form adopted for $\psi (\tau )$. The two parameters $T$ and $\mu $ should be
interpreted as some short- and long- time properties, respectively.

In summary, the system $S$ is an ensemble of dichotomous signals $%
\xi_{S}(t)$. Each system of this ensemble is determined by a sequence of
waiting times $\{\tau_i\}$ fitting the renewal property
$Pr(\tau_i,\tau_j)=Pr(\tau_i)Pr(\tau_j)$.
This sequence of waiting times generates a series of collisions occurring at
times $\{t_n\}$ given by 
$t_n=\sum_{i=0}^{n}\tau_i$. 
Each fluctuation $\xi_{S}(t)$ is determined by the following prescription:
At times $t= t_{n}$ the variable $\xi_{S}(t)$ gets either the value $+1$ or $%
-1$ with equal probability $1/2$ and keeps it till the next collision.


To demonstrate CME we assign the role of perturbation to another
NPR of the same kind. Hereafter we assign the subscript $S$ to indicate the
system properties. The system's waiting-time distribution is therefore $\psi
_{S}(t)$ of the form (\ref{theoretical}), with parameters $T_{S}$ and $\mu
_{S}$. The subscript $P$ is analogously adopted for the perturbation. The
perturbation is a dichotomous fluctuations $\xi _{P}(t)$, with the values $+1
$ and $-1$, with the statistics of collisions and jumps defined in the same
way as for the system fluctuations $\xi _{S}(t)$, thereby involving the
survival probability $\Psi _{P}(t)$ 
and the waiting time distribution $\psi _{P}(t)$. 
This perturbation acts on an ensemble of systems characterized by the
survival probability $\Psi _{S}(t)$ and the waiting time distribution $\psi
_{S}(t)$. This perturbation modulates the production rate of system jumps
with the following prescription. When the coupling between system and
perturbation is switched on, the occurrence of a system collision at time $t$
generates the state $|+\rangle $ with the probability $[1+\epsilon \xi
_{P}(t)]/2$ and the state $|-\rangle $ with probability $[1-\epsilon \xi
_{P}(t)]/2$. The parameter $\epsilon <1$ is a non-negative number defining
the interaction strength. If $\mu _{S}=\infty $, namely, when the systems
fits the Poisson condition, and $\xi _{P}=\cos (\omega t)$, this
prescription generates the well known phenomenon of stochastic resonance 
\cite{stochasticresonance}.

Due to the lack of a stationary correlation function, the response cannot be
expressed by means of the ordinary Green-Kubo theory \cite{green,thomas}.
Details on how to derive the linear response in this case can be found in
Ref. \cite{gianluca}, which leads to a theoretical prescription coincident
with the work of Refs. \cite{recentobservation}. Thus here we limit
ourselves to adopting the resulting exact expression that reads 
\begin{equation}
\langle \xi _{S}(t)\rangle =\epsilon \int_{0}^{t}dt^{\prime }P(t^{\prime
})\Psi _{S}(t-t^{\prime })\xi _{P}(t^{\prime }).  \label{key1}
\end{equation}
Note that, as in the case of ordinary stochastic resonance, we are making an
average on infinitely many responses to the same perturbation $\xi _{P}(t)$.
Here $P(t)$ denotes the rate of collision per unit
time, a property of the unperturbed system that remains unchanged even when
the coupling between system and perturbation is switched on. Using renewal
theory, it is straightforward to show that 
$P(t)=\sum_{n=1}\psi _{n}(t)$, 
where $\psi _{n}(t)$ is the probability density of undergoing the $n$-th
collision between times $t$ and $t+dt$, which is in turn related to $\psi
_{S}(t)$ through the iteration relation 
$\psi _{n}(t)=\int_{0}^{t}\psi _{n-1}(t^{\prime })\psi _{S}(t-t^{\prime })dt^{\prime}$. 
As a result, in the Laplace representation, 
\begin{equation}
\hat{P}(u)=\frac{\hat{\psi}(u)}{1-\hat{\psi}(u)}.  \label{rate}
\end{equation}
In the case $\mu _{S}<2$, the inverse Laplace transform of Eq. (\ref{rate})
yields, for $t\rightarrow \infty $, $P(t)\propto 1/t^{2-\mu _{s}}$,
indicating that the rate of collision production decreases as a function of
time, which is a different way of expressing the aging condition mentioned earlier.

CME implies that the perturbation is also an erratic signal. Thus,
the only possible way to make an analytical prediction rests on using
infinitely many realizations of the same perturbation signal $\xi _{P}(t)$
so as to produce a Gibbs ensemble of perturbations $P$ as well as a Gibbs
ensemble of systems $S$. Let us assume that all the perturbation signals $%
\xi _{P}(t)$ are properly prepared at time $t=0$ in the condition $\xi
_{P}(0)=1$, while the system is prepared in a condition with half the
systems in state $|+\rangle $ and half in the state $|-\rangle $, so that $%
\langle \xi _{S}(0)\rangle =0$. In practice, as an effect of this
prescription, we average Eq. (\ref{key1}) over all possible realizations of
the perturbation $\xi _{P}(t)$, which yields the double average $\langle
\langle \xi _{S}(t)\rangle \rangle $ in terms of the simple average $\langle
\xi _{P}(t)\rangle $. It is known \cite{gerardo} that 
$\langle \xi _{P}(t)\rangle =\Psi _{P}(t)$. 
Thus, due to the linear nature of Eq. (\ref{key1}) we obtain 
\begin{equation}
\langle \langle \xi _{S}(t)\rangle \rangle =\epsilon \int_{0}^{t}dt^{\prime
}P(t^{\prime })\Psi _{S}(t-t^{\prime })\Psi _{P}(t^{\prime }).  \label{key22}
\end{equation}


We are now properly equipped to make analytical predictions. Due to the time
convolution nature of Eq. (\ref{key22}) it is convenient to evaluate its
Laplace transform, and to consider the asymptotic limit $u\rightarrow 0$,
which by inverse Laplace transform yields information in the time asymptotic
limit $t\rightarrow \infty $. The calculations are done using the approach
illustrated in Ref. \cite{bruce} and yield, for $t\rightarrow \infty $,

\begin{equation}
\langle \langle \xi _{S}(t)\rangle \rangle =\epsilon \left[ A_{S}\left( 
\frac{T_{S}}{t}\right) ^{\mu _{S}-1}+A_{P}\left( \frac{T_{P}}{t}\right)
^{\mu _{P}-1}\right] .  \label{fenomeno}
\end{equation}
We see that the two terms of the r. h. s. of (\ref{fenomeno}) are
respectively proportional to the asymptotic tails of $\Psi _{S}$ and $\Psi
_{P}$ with amplitudes given by the analytical expressions:

\begin{eqnarray}
\label{AS}
A_{S} &=&-\frac{\Gamma (\mu _{P}-\mu _{S})}{\Gamma (\mu _{P})\Gamma (1-\mu
_{S})},\mbox{       and}  \\
\label{AP} 
A_{P} &=&\frac{\Gamma (1-\mu _{P}+\mu _{S})}{\Gamma (2-\mu _{P})\Gamma (\mu
_{S})}-\frac{\Gamma (\mu _{S}-\mu _{P})}{\Gamma (\mu _{S})\Gamma (1-\mu +P)}.
\end{eqnarray}
$A_{S}$ and $A_{P}$ have opposite signs, with the former being positive for $\mu _{S}<\mu _{P}$ and negative for $\mu _{S}>\mu _{P}$. In both cases the
dominating term is the one with the positive pre-factor, signaling that the
system is able to inherit the statistical properties of the perturbing
system when $\mu _{S}>\mu _{P}$, while maintaining its own statistical
properties when $\mu _{S}<\mu _{P}$. At $\mu _{S}=\mu _{P}$ both terms
diverge and this is due, as we shall see by means of a numerical
calculation, to the critical slowing down corresponding to the matching
condition.

To shed light on the physical meaning of this analytical result, it is
convenient to compare Eq.(\ref{fenomeno}) to the exact solution of Eq. (\ref
{key22}) in the case when the system satisfies the Poisson condition with rate 
$g_{S}$ and the perturbation is another Poisson system with rate $g_{P}$. In
this case 
\begin{equation}
\langle \langle \xi _{S}(t)\rangle \rangle =A_{S}\Psi _{S}(t)+A_{P}\Psi
_{P}(t),  \label{fittingfunction}
\end{equation}
$A_{S}={g_{S}}/{(g_{P}-g_{S})}$, $A_{P}=-A_{S}$, $\Psi _{S}(t)=\exp (-g_{S}t)
$ and $\Psi _{P}(t)=\exp (-g_{P}t)$. We see that the excitation-relaxation
process corresponds to the superposition of two exponential contributions
with coefficients of equal amplitude and opposite sign, with a diverging
maximum at $g_{P}=g_{S}$. This example, closely connected to the aperiodic
stochastic resonance \cite{aperiodic}, shows that the maximum response of
the system $S$ to the excitation $\xi _{P}(t)$ is obtained when the rate of
the perturbation is identical to the rate of the system. This suggests the
adoption of Eq. (\ref{fittingfunction}) as a fitting function of the Monte
Carlo calculation that we use to simulate CME. In this latter case 
$\Psi _{S}(t)$ and $\Psi _{P}(t)$ are given by Eq. (\ref{correlationfunction}), 
with subscripts $S$ and $P$, respectively. 
Using Eq. (\ref{fittingfunction}), we fit the values of $A_{S}$ and $A_{P}$
to the Monte Carlo data. CME is adequately described by Fig. 1,
which is obtained by keeping $\mu _{S}=1.6$ and moving $\mu _{P}$ from $\mu
_{P}=1.1$ to $\mu _{S}=1.95$. We plot the intensity of the slowest response
contribution as a function of $\mu _{P}$. We see that there is an impressive
similarity with the Poisson resonance phenomenon. The maximum response is
obtained by realizing the matching condition $\mu _{P}=\mu _{S}$, in analogy with
the resonance condition $g_{P}=g_{S}$. Two fitting examples, $\mu _{S}=1.35$
and $\mu _{S}=1.85$ are shown in the inset of Fig. 1. 

\begin{figure}[h]
\begin{center}
\vspace{-1.3cm}
\includegraphics[scale = 0.87]{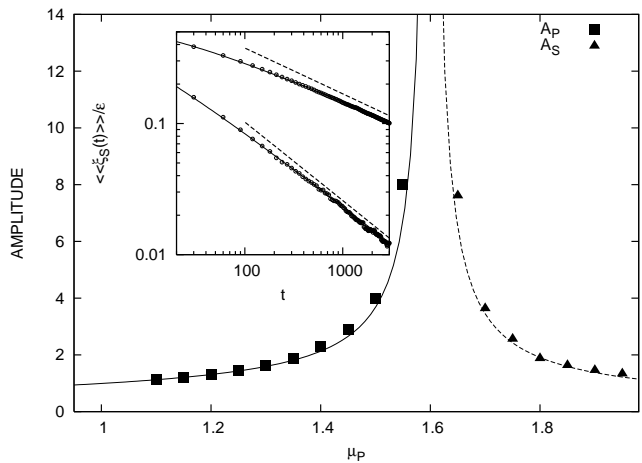}
\end{center}

\caption{Inset: fitting of Eq. (\ref{fittingfunction}) (solid lines) to Monte Carlo
data (open circles) using $T_S=T_P=1$, $\mu_S=1.6$ with $\mu_P=1.35$ (upper) and $\mu_P=1.85$ (lower). Dashed lines are the asymptotic dominant term in Eq. (\ref{fittingfunction}). Our Monte Carlo used $10^7$ system-perturbation pairs. 
Main figure: Amplitudes $A_P$ (squares), $A_S$
(triangles),
 Eqs. (\protect\ref{AP}) (solid line) and (\protect\ref{AS}) (dashed line)
  as a function of $\mu_P$, with $\mu_S=1.6$.}
\label{fit}
\end{figure}



The CME is a general property of a large class of complex systems and
is not limited to the intermittent fluorescence used to motivate our discussion.
For example, 
other NPRs are given by return times 
for a random walk, either in regular lattices or in a complex
network, regardless of whether the diffusion process is normal or
Fractional Brownian Motion \cite{arkady}. In general, events are generated
by a slow motion, which reflects the creation and the subsequent dynamics of
self-organized structures. Events either break these collective modes or
reflect the system entering a regime with high rates of entropy increase. This
can be, for instance, a passage in a chaotic sea, the effect of a
coin-tossing or a catastrophic "quake" in the system structure \cite{sibani}. 
Whatever the case, the system memory is reset, while the non-Poisson
behavior of $\psi(\tau)$,  with uncorrelated waiting times, 
is responsible for long-range correlations in the
observable sequence. 



\emph{Acknowledgments} We thankfully acknowledge Welch and ARO for partial
support through Grants B-1577 and Grant N. W911NF-05-1-0059, respectively.

\end{document}